\documentclass[aip,amsmath,amssymb,reprint,]{revtex4-1}
\usepackage{graphicx}% Include figure files
\usepackage{dcolumn}% Align table columns on decimal point
\usepackage{bm}% bold math
\usepackage[utf8]{inputenc}
\usepackage[T1]{fontenc}
\usepackage{mathptmx}
\usepackage{etoolbox}
\usepackage[usenames,dvipsnames]{color}
\usepackage[most]{tcolorbox}
\usepackage{multirow}
\usepackage{gensymb}
\usepackage[normalem]{ulem}
\usepackage[most]{tcolorbox}
\usepackage[colorlinks, linkcolor=blue,anchorcolor=blue,citecolor=blue,urlcolor=blue]{hyperref}
\usepackage{amssymb}
\usepackage{pifont}
\usepackage{physics}
\usepackage{natbib}
\usepackage{dsfont}
\usepackage{CJK}

\makeatletter
\def\@email#1#2{%
 \endgroup
 \patchcmd{\titleblock@produce}
  {\frontmatter@RRAPformat}
  {\frontmatter@RRAPformat{\produce@RRAP{*#1\href{mailto:#2}{#2}}}\frontmatter@RRAPformat}
  {}{}
}%
\makeatother
\begin{document}

\preprint{AIP/123-QED}
\begin{CJK*}{UTF8}{}

\title{Fast wide-field quantum sensor based on solid-state spins integrated with a SPAD array}

\author{Guoqing Wang \CJKfamily{gbsn}(王国庆)}
\thanks{These authors contributed equally.}
\email[]{gq\_wang@mit.edu}
\affiliation{
   Research Laboratory of Electronics, Massachusetts Institute of Technology, Cambridge, MA 02139, USA}
\affiliation{
   Department of Nuclear Science and Engineering, Massachusetts Institute of Technology, Cambridge, MA 02139, USA}
   
\author{Francesca Madonini}
 %\altaffiliation[Also at ]{Physics Department, XYZ University.}
\thanks{These authors contributed equally.}
\affiliation{
   Research Laboratory of Electronics, Massachusetts Institute of Technology, Cambridge, MA 02139, USA}
\affiliation{
   Dipartimento di Elettronica, Informazione e Bioingegneria (DEIB), Politecnico di Milano, Piazza Leonardo da Vinci 32, 20133, Milano, Italy}   
   
\author{Boning Li}
\thanks{These authors contributed equally.}
\affiliation{
   Research Laboratory of Electronics, Massachusetts Institute of Technology, Cambridge, MA 02139, USA}
\affiliation{Department of Physics, Massachusetts Institute of Technology, Cambridge, MA 02139, USA}

\author{Changhao Li}
\thanks{Present address: JPMorgan Chase, New York, NY, USA}
\affiliation{
   Research Laboratory of Electronics, Massachusetts Institute of Technology, Cambridge, MA 02139, USA}
\affiliation{
   Department of Nuclear Science and Engineering, Massachusetts Institute of Technology, Cambridge, MA 02139, USA}
   
\author{Jinggang Xiang}
\affiliation{
   Research Laboratory of Electronics, Massachusetts Institute of Technology, Cambridge, MA 02139, USA}
\affiliation{Department of Physics, Massachusetts Institute of Technology, Cambridge, MA 02139, USA}

\author{Federica Villa}
\affiliation{
   Dipartimento di Elettronica, Informazione e Bioingegneria (DEIB), Politecnico di Milano, Piazza Leonardo da Vinci 32, 20133, Milano, Italy}  

\author{Paola Cappellaro}
\email[]{pcappell@mit.edu}
\affiliation{
   Research Laboratory of Electronics, Massachusetts Institute of Technology, Cambridge, MA 02139, USA}
\affiliation{
   Department of Nuclear Science and Engineering, Massachusetts Institute of Technology, Cambridge, MA 02139, USA}
\affiliation{Department of Physics, Massachusetts Institute of Technology, Cambridge, MA 02139, USA}

\date{\today}

\begin{abstract}
Achieving fast, sensitive, and parallel measurement  of a large number of quantum particles is an essential task in building large-scale quantum platforms for different quantum information processing applications such as sensing, computation, simulation, and communication. Current quantum platforms in experimental atomic and optical physics based on CMOS sensors and CCD cameras are limited by either low sensitivity or slow operational speed. Here we integrate an array of single-photon avalanche diodes with solid-state spin defects in diamond to build a fast wide-field quantum sensor, achieving a frame rate up to 100~kHz. We present the design of the experimental setup to perform spatially resolved  imaging of quantum systems. A few exemplary applications, including sensing DC and AC magnetic fields, temperature, strain, local spin density, and charge dynamics, are experimentally demonstrated using an NV ensemble diamond sample. The developed photon detection array is broadly applicable to other platforms such as atom arrays trapped in optical tweezers, optical lattices, donors in silicon, and rare earth ions in solids.
\end{abstract}

\maketitle
\end{CJK*}

\section{Introduction}
The second quantum revolution describes all those new technologies that are enabled by the use of quantum mechanics, not only to describe the physical world, but to address, control, and detect individual quantum systems. For example, observing the behavior of individual atoms or photons enables the realization of phenomena such as quantum superposition and entanglement~\cite{horodecki_quantum_2009}. Nowadays, innovation is pushed in four main directions: quantum communication~\cite{gisin_quantum_2007}, to transmit data more securely; quantum simulation~\cite{georgescu_quantum_2014}, to reproduce physical dynamics in well-controlled systems; quantum computation~\cite{NielsenChuang_book}, to speed up computations; and quantum sensing~\cite{degen_quantum_2017}, to improve measurement performance. %Developing real-world applications is still challenging, mainly because quantum systems require control that preserves their quantum properties from the noise, decoherence, and alterations that each observation generates~\cite{suter_colloquium_2016,lidar_brun_2013}. 
Despite great advances, it is still challenging to achieve high-fidelity control of scalable and low-noise quantum systems~\cite{suter_colloquium_2016,lidar_brun_2013}.

Just in the field of measurement, the need to address single quantum particles, rather than a macroscopic sample, requires detectors with high sensitivity (e.g. single-photon sensitivity for optical measurements)~\cite{zhang_material_2020-1,saffman_quantum_2010}. Moreover, to scale up quantum systems for more powerful quantum information processing, it is desirable to achieve the simultaneous measurement of a larger number of quantum particles. Exemplary application scenarios include wide-field quantum sensing with solid-state spins~\cite{scholten_widefield_2021}, large-scale quantum computation with Rydberg atom arrays~\cite{saffman_quantum_2010}, and quantum simulation with optical lattices~\cite{gross_quantum_2017}. In addition, quantum error correction requires fast  and high-fidelity measurements~\cite{xu_fast_2021} to enable feedback on the quantum platforms based on the detected results. 

Single-Photon Avalanche Diodes (SPADs) integrated in CMOS technologies are  widely used today to detect single photons, providing good overall performance with the typical advantages of microelectronic technologies, such as reliability, robustness, and compactness. Their relatively high Photo-Detection Efficiency (PDE), along with the digital-like output (every time at least a photon is absorbed) and the absence of readout noise, makes them an arguably suitable choice for applications requiring single-photon counting or Time-Of-Flight (TOF) evaluation. In particular, SPADs are well suited to the realization of large-format arrays for imaging applications since they allow single-photon sensitivity together with high frame rates and spatial resolution~\cite{cusini_historical_2022}. CMOS Active Pixel Sensors (APS) are high-speed devices, but they do not perform internal amplification, thus suffering from low sensitivity. Electron-Multiplying Charged Coupled Devices (EMCCD) and Intensified CCDs (ICCD), although having high sensitivity and possibly millions of pixels, are bulkier, slower, and more expensive than SPAD arrays~\cite{madonini_single_2021}.
%I can add references for SPADs

In this work, we present the development of an advanced detection system based on the MPD-SPC3 SPAD camera~\cite{SPC3,ndagano_imaging_2020}, targeting light extraction from single photon emitters that can be used as quantum sensors. In comparison to state-of-the-art photon detection such as single-pixel commercial photodiodes, or spatially resolved CCD cameras in case of bulk measurements, our work is the first integration of an array of SPADs with an ensemble of Nitrogen-Vacancy (NV) centers in diamond, providing micrometer-scale spatial resolution and fast measurement rate (100~kHz). We build an optical microscope setup with flexible magnification factors to map the top layer of a bulk diamond to the $64\times 32$ array of SPAD pixels. We achieve efficient transmission and processing of the fast-generating ($>10$~Mbps) SPAD data. We benchmark our new system by demonstrating proof-of-principle examples of sensing AC and DC magnetic field, temperature, strain, local spin density, and charge dynamics with the NV ensemble. Its quantum applications for broader solid-state and atom array platforms are discussed in the outlook.

\begin{figure}
\includegraphics[width=0.5\textwidth]{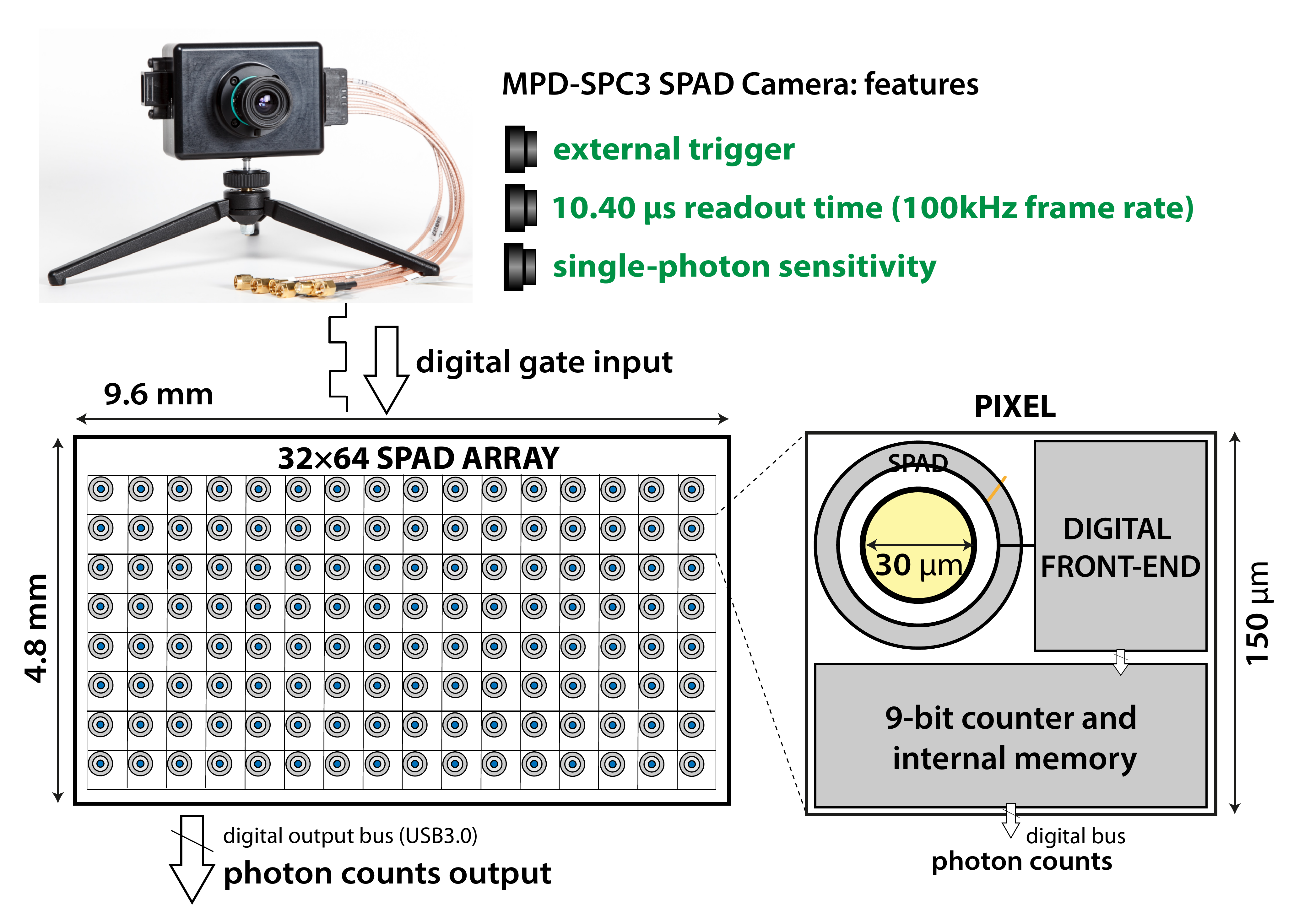}
\caption{\label{fig0:Camera} \textbf{MPD-SPC3 SPAD Camera.}}
\end{figure}
\section{MPD-SPC3 SPAD camera}
A SPAD is a p-n junction that is reverse-biased well above its breakdown voltage ~\cite{cusini_historical_2022-1}. Under this operating condition, the absorption of a single photon generates an electron-hole pair, which is accelerated by the high
electric field across the junction. The energy of the charge carriers is eventually sufficient to trigger a macroscopic avalanche current of few milliamperes through the device. The SPAD front-end electronics, namely
the quenching circuit, can be integrated on the same silicon chip. The quenching circuit senses the avalanche current, quenches it and resets the SPAD to its initial state. The avalanche is quenched by decreasing the voltage across the SPAD junction below breakdown for an adjustable time (hold-off), that ranges from few nanoseconds to few hundreds of nanoseconds. The SPAD bias voltage is then restored to the initial state, and thus  it is ready to detect the next photon. The total time required to restore the initial state of the SPAD after the detection of a photon is named dead time. During the dead time, no photons can be detected. Due to the dead time, the SPAD has a linear working range (i.e., the number of detection events within a defined time period (T) depends linearly on the illumination intensity) only at low or moderate photon fluxes. At large photon fluxes, the number of detected photons deviates from linearity and saturates to a constant value, which is the reciprocal of the dead time. 
The measurement of light signals by a SPAD has several advantages concerning the signal-to-noise ratio. No analog measurement of voltage or current is needed, since the detector acts like a digital "Geiger-like" counter. It follows that no electronic noise is added by analog-to-digital converters or amplifiers while measuring the signal. Additionally, the detector is less sensitive to electromagnetic interference or electrical noise generated by external equipment, differently from CCDs. 

The "MPD-SPC3"~\cite{SPC3} as shown in Fig.~\ref{fig0:Camera} is a multipurpose single-photon counting image sensor, based on a 0.35 $\mu$m $64\times32$ SPAD pixel array. Each smart pixel comprises a very low noise (100 cps) round SPAD with 30 $\mu$m active area diameter, a low area occupation analog quenching circuit, and digital processing electronics. The quenching front-end provides a digital pulse to the following counter block every time at least a photon impinges on it. Three 9-bit LFSR (Linear Feedback Shift Register) counters are employed to deliver two-dimensional intensity information through photon counting in either free-running (down to 10-$\mu$s integration time) or time-gated mode. Free-running is the default operation and consists in continuously counting all SPAD pulses without any temporal filtering. Conversely, in time-gated mode, each of the three integrated binary counters is enabled only inside a user-defined temporal window (gate signal). The LFSR topology grants high-speed operation and low area occupation. Three in-pixel memory registers store the content of the respective counter, simultaneously for the whole array, thus allowing global shutter operation. The array readout time is proportional to the number of counters in use. In case a single counter per pixel is used,  this time is equal to the time needed to read the 2048 pixels, which is 10.40-$\mu$s. The devised pixel has a pitch of 150 $\mu$m and a fill factor of 3.14\%. For imaging applications, microlenses can improve the fill-factor to about 78\%. 
% PLEASE CHECK THE ABOVE ##
The SPAD camera features  high Photon-Detection Probability (PDP) in the visible and near-UV spectral region with a peak of almost 50\% at 410 nm. The minimum dead time is 50 ns, ensuring a maximum count rate of 20 Mcps. The dark count rate, i.e., the rate of avalanches not caused by photon detection and mainly related to thermal carrier generation, is very low even at room temperature (100 cps per pixel). 

\section{Sensor setup design with flexible control and readout options}
\begin{figure*}
\includegraphics[width=\textwidth]{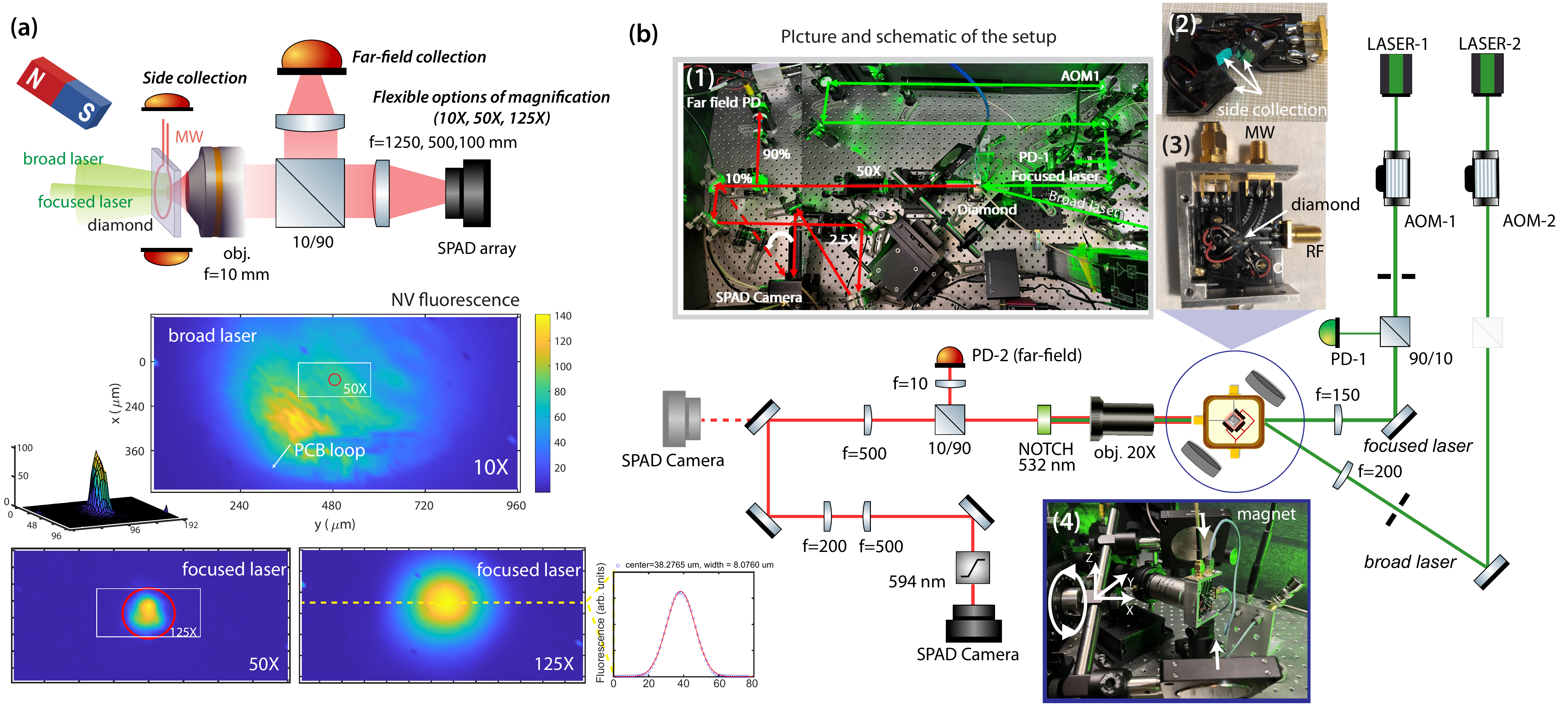}
\caption{\label{fig1:setup} \textbf{Experimental setup design.} (a) Simplified schematic of the setup design. The NV fluorescence measured under different magnification factors and laser settings are plotted. The focused laser beam size is characterized by plotting a line cut of the NV fluorescence imaging (inset). (b) Setup design and pictures. Two laser beams with flexible sizes on the diamond are gated by two AOMs. The maximum magnification factor of 125 is realized by two magnification steps (50$\times$ and 2.5$\times$). The second step can be bypassed when only small magnification factors are needed. }
\end{figure*}
NV centers in diamond have been one of the most promising solid-state platforms for quantum sensing~\cite{degen_quantum_2017}, with optical initialization and readout of the spin state, room-temperature operational condition, and nanoscale spatial resolution. Its capabilities of sensing electromagnetic signals~\cite{taylor_high-sensitivity_2008,maze_nanoscale_2008,dolde_electric-field_2011-2}, temperature~\cite{kucsko_nanometre-scale_2013}, strain/stress~\cite{hsieh_imaging_2019}, spin density~\cite{li_determination_2021}, charge environment~\cite{dhomkar_charge_2018} have been demonstrated in recent years. To further improve the readout efficiency and perform simultaneous measurement of different spatial locations, wide-field quantum sensing techniques have also been developed~\cite{scholten_widefield_2021}, where a microscope objective is used to map out the spatial information, and a camera with multiple pixels is used for imaging a large number of pixels. In this section, we introduce the details of our setup design, highlighting flexible control and readout options tailored for different application scenarios.

%The NV center in diamond is one of the many possible solid-state systems used in quantum magnetometry, which exploits the non-classical properties of such devices to provide highly sensitive measurements of magnetic fields. The NV center, indeed, contains a magnetically-sensitive electronic ground state structure (made of a spin triplet) that can be initialized, coherently controlled using microwaves, and optically readout. The designed setup can roughly be divided into eight subsystems.

\subsection{Sample mounting, magnetic field, and microwave control}
The foundational element in an NV sensing experiment is the diamond host itself. Tailoring the diamond growth recipe is vital to obtain the desired amount of NV defects while controlling the concentration of other paramagnetic centers. The diamond itself is then mounted to allow for easy access by a 532-nm excitation laser, and for collection of the resulting NV fluorescence. The diamond mount is designed to minimize thermal gradients, vibrational noise, and additional strain induced by any adhesive applied to the diamond. In our setup, the diamond is mounted over a hole of diameter $\sim1$~mm on a printed circuit board (PCB) with the conducting surface layer finished by immersion gold. The thickness of the PCB is chosen as 0.4~mm (much thinner than the typical PCB thickness of around 1.6~mm) to allow more photon collection via the objective mounted on the backside of the PCB board. 

%\subsection{Bias Magnetic Field}
For NV experiments not operating at zero-field conditions, a bias magnetic field is applied to break the degeneracy of the two spin states $\ket{m_s=\pm1}$ and  spectrally distinguish the $\ket{m_s=0}\leftrightarrow\ket{m_s=+1}$ and $\ket{m_s=0}\leftrightarrow\ket{m_s=-1}$ transitions. We use a pair of permanent magnets mounted on a stage with $x-y-z$ translation freedom and rotation freedom. The distance between the two magnets can also be freely adjusted (shown in the inset (4) of Fig.~\ref{fig1:setup}(b)).

%\subsection{Microwave delivery}       
One typical NV sensing method is to scan the frequency of an applied microwave field to find the NV resonance condition, which changes depending on the magnetic field to be detected. Applying resonant microwaves then allows the realization of more complicated and precise sensing protocols. Typically, a wire loop is constructed by shorting a coaxial cable, which is placed near the NV interrogation region. A microwave generator then generates a field, which is amplified and sent to the wire to drive the NV transitions. Other microwave delivery schemes include using a waveguide fabricated directly on the surface of the diamond, which however lacks convenience for sample change. In our case, the PCB holding the diamond also hosts the microwave wire loop, formed from the end of a shorted coaxial cable. The diamond is mounted such that the NV layer (usually near one of the surfaces of the diamond) is facing toward the loop (actually touching the loop) to maximize the microwave power delivered to the NV ensemble. 

\subsection{Multiple laser beam settings}        
Optical excitation is necessary to polarize the NV center to the $\ket{m_s=0}$ state~\cite{schirhagl_nitrogen-vacancy_2014}. For typical single-NV or small-ensemble volume experiments based on a confocal setup design, a microscope objective is used to both apply the optical excitation beam and detect the fluorescence signal. When applying the excitation beam, the diamond is placed at the output of the microscope objective, allowing only NVs in the diffraction-limited spot (usually 200~nm to 1~$\mu$m, depending on the numerical aperture and laser wavelength)  to be optically addressed by the laser. For wide-field imaging applications, we use the microscope objective only for fluorescence detection and apply a broad laser beam from the opposite side.

To allow higher experiment flexibility such as the sensing of charge dynamics, we include two different green (532 nm) laser beams. Indeed, in standard setups, the optical readout of the NVs is performed with the same laser used for optical polarization. The use of two lasers allows the excitation and detection to be more flexible. For example, by creating a beam with a narrower focus and  a higher power density and a second, broader beam with a smaller power density, we can explore charge dynamics by tuning the local electron and hole density using the ionization and recombination process of the NV center\cite{wang_manipulating_2023}. The focused beam can be used for local charge states and electron/hole density control, while the broad beam can be used  for a homogeneous readout on the entire area. Besides focused-broad beam settings, more scenarios can be easily implemented such as making the two beams both focused on two different locations.

We use beamsplitters together with half-wave plates to generate the two laser beams from the same laser source (SPROUT-G, 5~W). Each of them is independently gated by an Acousto-Optic Modulator (AOM) in combination with an iris. The gating signals are provided by a pulse blaster (PulseBlasterESR-PRO 500) such that the laser can be turned on and off in experimental sequences. The optical paths shown in Fig.~\ref{fig1:setup}(b) are designed to have LASER-1 very focused (30 $\mu$m-diameter) and LASER-2 broader (>200 $\mu$m-diameter).

\subsection{Multiple detection options}
\subsubsection{Side collection and far-field collection}
Due to total internal reflection, for a bulk diamond plate with a thin thickness in the $z$ direction, more fluorescence light actually emerges from the side edges~\cite{le_sage_efficient_2012}. We devise a separate piece of PCB based on the idea of Ref.~\cite{le_sage_efficient_2012}, where three photodiodes (area 2~mm~$\times$~2~mm each, with attached a thin customized long-pass filter  to block the excitation light) are placed on the side of the diamond (shown in the inset of Fig.~\ref{fig1:setup}(b)). The instantaneous current of the photodiode (Hamamatsu S8729) is sampled and converted to voltage signal by a variable gain high-speed (200~MHz) current converter (DHPCA-100), which is then measured by a National Instrument data acquisition card (NI PCI-6281) gated by the pulse blaster. With a high signal-to-noise ratio and data acquisition efficiency, the side collection detection option was mostly used in our previous work where no spatial information was required~\cite{wang_coherence_2020-1,wang_characterizing_2022}

Even though separate pieces of PCB are used for delivering microwaves and performing data collection as shown in the insets (2,3) of Fig.~\ref{fig1:setup}(b), the side collection suffers from  circuit interference due to microwave driving and laser heating. Thus, we also use a microscope objective to collect fluorescence from the backside of the PCB. The fluorescence signal is then sent to a variable gain photodiode (PBA36A). Both  collection options can be operated simultaneously, providing better fluorescence collection efficiency.

%\subsubsection{Laser noise collection}        
%The initial setup didn't include an objective. When it was added for magnification purposes, the signal directed to the SPAD camera became excessively high. Therefore, not to waste light signal, we added two additional detection paths. 
To cancel the noise due to laser intensity fluctuation, we add a 10/90 beamsplitter to sample a fraction of the intensity of either of the two laser sources. The signal is then collected by a separate photodiode (PBA36A), which is useful to characterize and compensate for the laser noise. When far-field detection is used for data collection, we can use a balanced photodiode (such as PDB230A) to perform the laser noise cancellation with even better performance. %Indeed, especially for long ionization/recovery measurements, it is important to subtract laser instability contribution. 

\subsubsection{SPAD array with multiple magnification factors} 
The wide-field fluorescence collection with MPD-SPC3 camera shares the same objective (N20X-PF - 20X Nikon Plan Fluorite Imaging Objective, 0.50 NA, 2.1 mm, 10 mm effective focal length)) as the far-field collection path. 
%The NV ensemble is excited with a 30 \textmu~m laser spot and a $\sim$1 \textmu~m resolution had to be ensured with the SPC3 camera (150 \textmu~m pixel pitch) in order to investigate potential temperature gradients and magnetic field inhomogeneities with such spatial resolution. Indeed, high-throughput measurements of spatial inhomogeneities allow a fast pre-screening of the diamond before quantum physics experiments. 
To match the SPC3 camera's $150\mu$m pixel size with the desired spatial resolution $\sim1\mu$m, and ensure the off-axis light can be well captured by the lenses, we design a two-stage magnification process to achieve the magnification factor of 125. As depicted in Fig.~\ref{fig1:setup}(b), a first 50$\times$ factor is reached with an achromatic lens with a 500~mm focal length. A second 2.5$\times$ magnification factor is obtained with a couple of achromatic lenses with focal lengths of 100~mm and 250~mm. As a standard rule, mirrors were used to gain centimeter distances on the limited optical table. In particular, flip mounts were exploited for mirrors in order to potentially bypass the second magnification and keep only the first 50$\times$ factor indicated by the white arrow in the insets (1) of Fig.~\ref{fig1:setup}(b). Moreover, by changing the second lens focal length, and consistently moving the SPAD camera, it is straightforward to reach a different magnification value (such as 10$\times$, employed in the experiments described below). Such interchangeable optics ensures a compact and flexible sensor for spatially resolved images of NV quantum sensor. In practical experiments, high-throughput measurements allow a fast pre-screening of the diamond in a much larger spatial range with a smaller magnification factor before performing finer experiments with larger magnification factors.

The diamond emits red fluorescence but also a large amount of the green laser passes through it. A notch filter (532 nm and 1064 nm) is placed after the objective to filter out the green light. Moreover, a long-pass filter (594 nm) on the SPAD camera eliminates the remaining green light in favor of red fluorescence.

%\subsection{Data processing}      
%With a second beam-splitter, the emitted red fluorescence is divided in two paths, one directed to the SPAD camera, as already described, and the other to an additional commercial photodiode, referred to as ``far-field photodiode", which plays the role of a total ensemble detector which the SPAD camera can be compared to. Both photodiodes' gain and bandwidth were studied to provide a synchronous and comparable measurement in respect to the SPAD camera. 
The integration time of the SPAD camera is set by its control interface on the experimental computer, while the timing of the data acquisition is gated by the same pulse blaster, which is programmed in each experiment to synchronize and trigger different electronic components including laser excitation (through AOMs), microwave radiation, and photodiodes' measurement (through the data acquisition card). The quickly generated SPAD data is transmitted to the computer through a USB 3.0 port. Efficient data processing, especially for data size of tens of gigabytes, can be achieved with the memory-mapping method based on MATLAB. %(codes are available per reasonable request to the corresponding authors).

%\gw{Boning, it would be great if you could help add some description related to Figure 1 where we show different magnifications using the NV fluorescence measurement under different laser options. Could you also write the caption for Fig 1? I will do the other figures.}

\section{Applications in quantum sensing}
\subsection{Microwave imaging by Rabi magnetometry}
\begin{figure}
\includegraphics[width=0.5\textwidth]{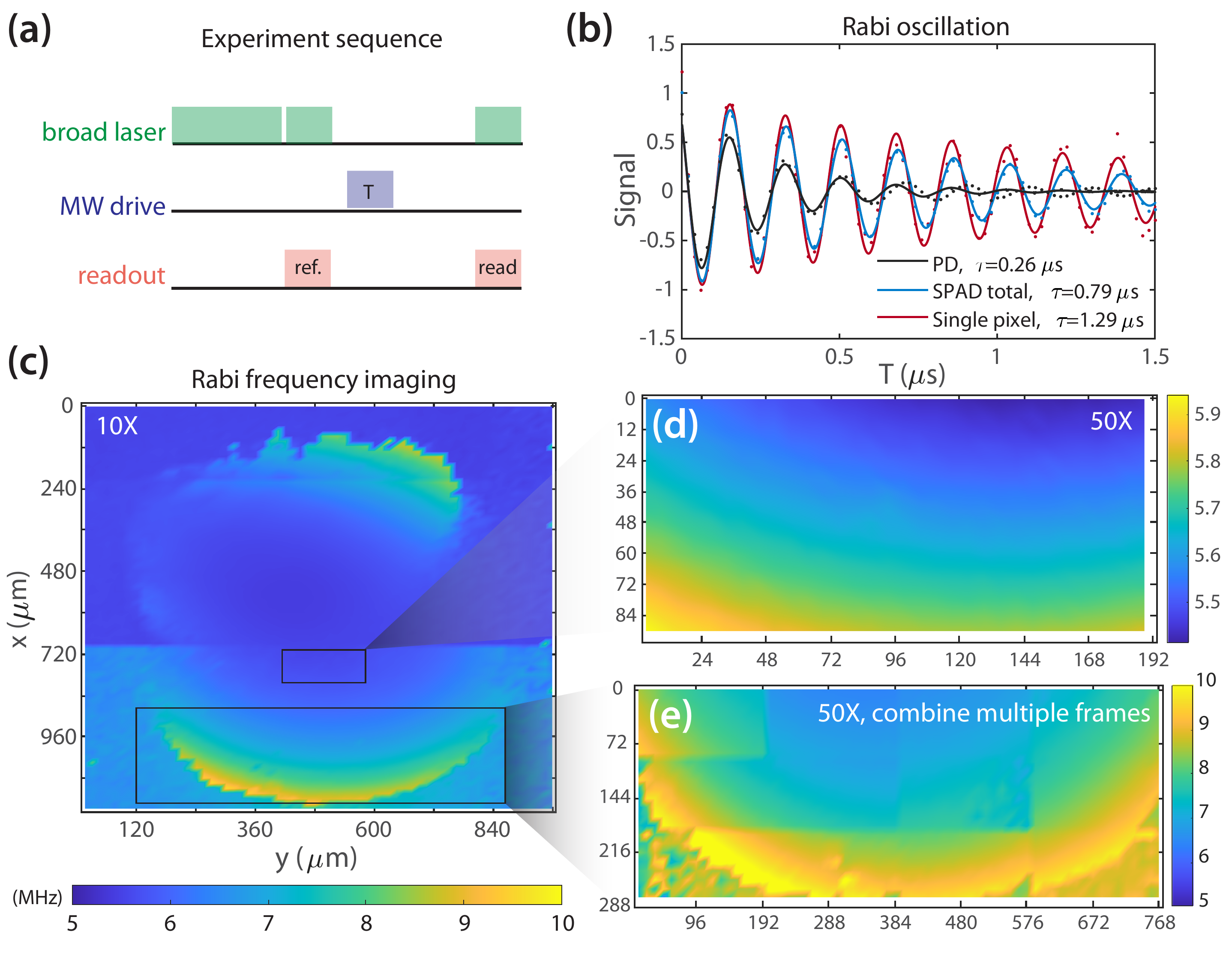}
\caption{\label{fig2:Rabi} \textbf{Microwave imaging.} (a) Experimental sequence for Rabi oscillation measurements. (b) Comparison of Rabi oscillation data collected by the far-field photodiode, individual and total pixels of the SPAD camera under the 50$\times$ magnification option (shown in d). (c) Imaging of Rabi frequency using the 10$\times$ magnification option of the optical path. Three frames with different $x$ stage positions are combined together. (d) A single frame Rabi frequency imaging with the 50$\times$ magnification option. (e) Rabi imaging with the 50$\times$ magnification option by combining 12 frames. }
\end{figure}

To sense an oscillating magnetic field with a frequency $\omega_s$, one can use the static bias field to tune the NV single quantum (SQ) transition frequency $\omega_0=D\pm\gamma_e B_z$ to this frequency (here $\gamma_e=(2\pi)2.8$~MHz/G is the gyromagnetic ratio, $D=(2\pi)2.87$~GHz is the zero-field splitting.) When the resonance condition is satisfied $\omega_0=\omega_s$, the NV center, starting from an initial state $\ket{0}$,  undergoes coherent oscillations between $\ket{0}$ and $\ket{-1}$ (or $\ket{+1}$) with a frequency $\Omega_s=\gamma_eB_s/\sqrt{2}$, called  Rabi frequency. Experimentally, we monitor the time-dependent population  $P_{\ket{0}}=(1+\cos(\Omega_s T))/2$, from which we obtain the Rabi frequency and then the amplitude $B_s$ of the target signal (Fig.~\ref{fig2:Rabi}(a))~\cite{alsid_solid-state_2022}.

As a demonstration of the flexibility of the designed setup, we use Rabi magnetometry to map the microwave field amplitude generated by the loop structure on our PCB. We use the broad laser beam to address a large area of the diamond. The measured Rabi oscillation with different collection options is shown in Fig.~\ref{fig2:Rabi}(b). By fitting the oscillation signal to a decaying sinusoidal $S(T)=c_0+c\cos(\omega_s T+\phi)e^{-T/\tau}$, we can obtain the coherence time $T_{2\rho}=\tau$ and Rabi frequency $\omega_s$. 
The single-pixel measurement has a longer coherence time (1.29~$\mu$s) than the average signal over all pixels (0.79~$\mu$s, both  measured under a 50$\times$ magnification factor). 
Similarly, the far-field photodiode measurement gives the shortest coherence time (only 0.26~$\mu$s) shown by the black data points in Fig.~\ref{fig2:Rabi}(b) because the signal is collected over a much larger spatial area (covering almost the whole area within the loop). In Fig.~\ref{fig2:Rabi}(c), we choose the 10$\times$ magnification option to image the microwave amplitude in a large field of view. We combine three frames measured under different $y$ stage positions and show the microwave amplitude distribution generated by the loop. The Rabi frequency near the proximity of the loop is close to  10~MHz, while in the central region it is quite homogeneous and is close to 5~MHz. In Fig.~\ref{fig2:Rabi}(d), we choose the 50$\times$ magnification option to have a finer measurement of the gradient in the central region of the loop structure, where we see the smaller distribution of the microwave amplitude between 5.4~MHz and 5.9~MHz. In Fig.~\ref{fig2:Rabi}(e), we combine twelve frames to measure a large region using the 50$\times$ magnification option, which images the gradient near the edge of the loop.

We note that the discontinuity of the image at the edge of each individual frame is due to different frequency detuning $\omega_0-\omega_s$ when measuring different frames. The effective Rabi frequency is then modified to $\Omega_{eff}=\sqrt{\Omega_s^2+(\omega_0-\omega_s)^2}$. These frequency mismatches may be induced by gradients of the magnetic field, temperature, or strain. The characterization of these gradients will be introduced shortly in the following sub-sections.

Our results demonstrate the capability of our setup in imaging microwaves with Rabi magnetometry. The coherence time $T_{2\rho}$ of Rabi oscillation strongly depends on the spatial range of the signal collection, which validates the decay mechanism due to spatial inhomogeneity of the microwave field proposed in our previous work~\cite{wang_coherence_2020-1}. In comparison to photodiode measurement giving rise to short coherence time, the SPAD measurement combines both benefits of long coherence time and spatially resolved information.

%Rabi frequency shows a dependence on the applied magnetic field, thus a spatially resolved Rabi measurement maps the magnetic wire loop used to deliver microwave pulses to the diamond sample. The results obtained with 125$\times$ magnification factor are shown in Figure (a), where five frames obtained by horizontally moving the objective are placed side by side. In the lower left corner, where an increase in Rabi frequency can be seen, you can sense the proximity of the magnetic wire loop. Figure (b) includes twelve images obtained with 50$\times$ (again by moving the objective), and the upper part of the loop can be seen (with related frequency increase). Finally, 7.5$\times$ factor allows to see the complete loop ($\sim$1 mm diameter) by combining two SPAD camera frames. The visible frequency jump between the top and bottom frames can probably be attributed to temperature effects created by the broad measurement laser. In the future, such laser beam could be made even broader to address uniform temperature over the camera frame. 

%\subsection{Magnetometry performance analysis} (better sensitivity? longer coherence time? better contrast? vs. SPAD saturation problem) \Red{this could be a separate paper targeting at Phys Rev Applied if we find something promising in sensitivity improvement, etc.}

\begin{figure*}
\includegraphics[width=\textwidth]{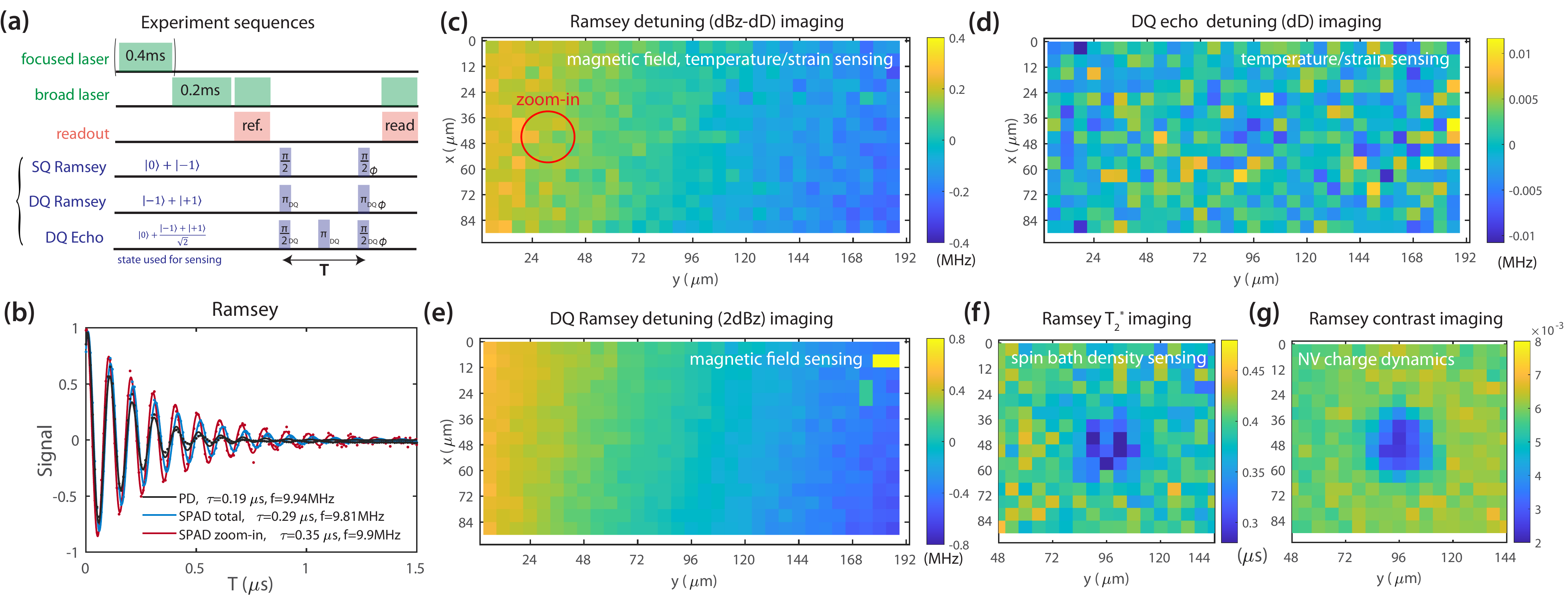}
\caption{\label{fig3:Ramsey} \textbf{Quantum sensing with wide-field imaging.} (a) Experimental sequence for SQ Ramsey, DQ Ramsey, DQ echo measurements. (b) SQ Ramsey experiments with different photon collection options and spatial ranges. The SPAD 64$\times$32 pixels are grouped to 32$\times$16 pixels to improve the signal-to-noise ratio (same for all the other data in this figure). (c) The detuning frequency imaging of SQ Ramsey experiments. The mean fitting uncertainty of the detuning value is 0.067~MHz. (d) The detuning frequency imaging of DQ echo experiments. The mean fitting uncertainty of the detuning value is 0.007~MHz. (e) The detuning frequency imaging of DQ Ramsey experiments. The abnormal spots in the upper right corner are due to the fitting failure in the data processing. The mean fitting uncertainty of the detuning value is 0.095~MHz. (f) The coherence time imaging of SQ Ramsey experiments with the focused laser beam illuminated as shown in sequence (a). The mean fitting uncertainty of the $T_2^*$ value is 0.045~$\mu$s. (g) The signal contrast imaging of SQ Ramsey experiments with the focused laser beam illuminated.  The mean fitting uncertainty of the contrast value is 0.0007. }
\end{figure*}

\subsection{Magnetic field imaging by Ramsey magnetometry}
To sense a static magnetic field, we can apply an initial $\pi/2$ pulse to prepare the spin state in the SQ superposition state $(\ket{0}+\ket{-1})/\sqrt{2}$. After an evolution time $T$, a relative phase $\varphi(T)=(dD-\gamma_edB_z)T$ is accumulated, before a final $\pi/2$ pulse  and projective optical measurement realizes  a population readout. Here $dD, dB_z$ are shifts of the zero-field splitting and the static magnetic field in different spatial locations with respect to their mean values calibrated by an experiment summing up the whole area. Usually, the final $\pi/2$ pulse is realized with a time-dependent phase $\phi=\nu T$ with respect to the first $\pi/2$ pulse, which adds an additional oscillation at a  frequency $\nu$ to the signal to better visualize the spin coherence behavior. The sequence is shown in Fig.~\ref{fig3:Ramsey}(a). When the temperature and strain are  constant and homogeneous,  SQ Ramsey magnetometry can be used to measure the static magnetic field $dB_z$.

We choose the 50$\times$ magnification option (same for all experiments below) to perform wide-field imaging of the magnetic field. The experimental results are shown in Fig.~\ref{fig3:Ramsey}(c), where a clear gradient along the $y$ direction can be seen. The deduced magnetic field change from $y=0$ to $y=192\mu$m is about $0.286$~G. Since the NV transition frequency inhomogeneity due to the magnetic field gradient is much smaller than the Rabi frequency, such a gradient would not affect the Rabi imaging measurement (Fig.~\ref{fig2:Rabi}) too much. Moreover, the characterized gradient may be used to improve microwave imaging.
%PC:  Maybe we can add here a comment that this would not affect the Rabi measurement too much
In Fig.~\ref{fig3:Ramsey}(b) we compare again the photodiode and SPAD measurements. The shortest Ramsey coherence time $T_2^*\sim0.19~\mu$s is observed in the photodiode measurement, while the average signal of the SPAD image gives 0.29~$\mu$s and the total signal of a small group of pixels shown by the red circle in (c) gives a longer 0.35~$\mu$s. We note that the gradient of temperature or strain that leads to variation in $D$ may also introduce a NV resonance frequency gradient. The magnetic field, temperature, and strain gradients, as well as the intrinsic spin bath with magnetic dipolar interactions, contribute to the decay of the Ramsey signal~\cite{bauch_decoherence_2020,bauch_ultralong_2018}.
% PC: we should here point out to some other mechanism beyond B-field inhomogeneity that lead to this decay. 
%PC: what's the minimum frequency shift we can measure? (that is, the map in Fig. 4 (c) (and similar) is  continuous in the magnetic field amplitude, while most probably we have a minimum shift that we can measure. We should explain. Also spatially, the map should be "pixelated" since that's what we measure, instead of having matlab doing an interpolation. 
% PC: the SQ image has much more variations than the DQ image, on a short spatial scale that is not easily compatible with a simple magnetic field gradient. Then, indeed, we might think that we do have e.g., strain effects

To further validate that the measured spatial gradient is given by the magnetic field $dB_z$ instead of the shift in zero-field splitting $dD$, we introduce the DQ Ramsey magnetometry, which uses an initial DQ superposition state $(\ket{+1}+\ket{-1})/\sqrt{2}$ prepared by simultaneously applying resonance microwave driving to both of the two SQ transitions. The accumulated relative phase at time $T$ is then $\varphi(T)=2\gamma_e dB_zT$, which is twice larger than the SQ Ramsey case when $dD=0$. The DQ Ramsey measurement is independent of the frequency shift in $D$, thus excluding the potential contribution from temperature, strain, or electric field, which mainly affects the NV transition frequency by changing the value of $D$. The experimental results are shown in Fig.~\ref{fig3:Ramsey}(e), which are consistent with the image in (c) (the gradient of the Ramsey detuning frequency in (e) is twice larger than the SQ measurement in (c)). 
%PC: what are the coherence times for the DQ? This looks really llike a nice linear gradient that you would expect to give rise to a sinc function when averaged. Is that consistent with the result? How do they compare to the SQ image? Also, for our own purposes: can we use the information that we have this y-gradient to actually align better the magnetic field?

\subsection{Temperature and strain sensing by double-quantum echo (or D-Ramsey)}
To sense the shift in zero-field splitting $D$, the magnetic field coupling term needs to be canceled by the double-quantum spin echo, which adds a double-quantum $\pi$ pulse in the middle of the Ramsey sequence. The method, also called D-Ramsey, has been used in measuring temperature~\cite{toyli_fluorescence_2013,neumann_high-precision_2013} and crystal strain~\cite{marshall_high-precision_2022}. Different from most previous work where the SQ superposition state is used as the initial state, here we use a superposition state $[\ket{0}+(\ket{+1}+\ket{-1})/\sqrt{2}]/\sqrt{2}$, which can be prepared by simultaneously applying resonance microwave driving to both of the two SQ transitions for a duration (half the duration of the initial pulse in DQ Ramsey, which prepares the state $\ket{+1}+\ket{-1}$), or by composite SQ pulses (a $\pi/2$ pulse applied to one SQ transition, before a $2\cos^{-1}\sqrt{2/3}$ pulse applied to the other SQ transition). The DQ $\pi$ pulse in the middle is composed of three SQ $\pi$ pulses applied to the $\ket{0}\leftrightarrow\ket{-1}$, $\ket{0}\leftrightarrow\ket{+1}$, $\ket{0}\leftrightarrow\ket{-1}$ transitions.

The measured frequency shift $dD$, plotted in Fig.~\ref{fig3:Ramsey}(d) shows quite small values (less than 0.01~MHz) in the whole image. The result demonstrates the homogeneous spatial distribution of both the temperature (<0.15~K) and strain (<1 ppm).
%PC: If I look at the images c-e-d I could think of c as being the magnetic field inhomogeneity of e + the train of d. This is probably not quantitative, but it could be at least mentioned qualitatively.

\subsection{Spin bath characterization and charge dynamics}
When analyzing the Ramsey decoherence time of NV centers within a single pixel (or a small group of pixels) such that the magnetic field, temperature, or strain inhomogeneity is small, the dominant noise is then given by the surrounding spins which are coupled to the NV spin through magnetic dipolar interaction. A simple dimensional analysis predicts that the decay rate $1/T_{2}^*$ should be proportional to the local spin density, as recently analyzed~\cite{li_determination_2021,stepanov_determination_2016}. %An analytical model has revealed that the decay rate $1/T_{2}^*$ is proportional to the local spin density~\cite{li_determination_2021,stepanov_determination_2016}.
%An analytical model has revealed that the decay rate $1/T_{2}^*$ is proportional to the local spin density~\cite{li_determination_2021,stepanov_determination_2016}.
Thus, characterizing the Ramsey decoherence provides a probe of the local spin density. To create a spin density gradient in our system, we use the focused beam with a large power density to quickly drive the cycling transition between the two charge states NV$^-$ and NV$^0$~\cite{jayakumar_optical_2016,wang_manipulating_2023}, which continuously pump electrons from the valence band to the conduction band. The generated electrons then combine with positively charged nitrogen defects (without spin) and convert them to neutrally charged nitrogen defects (P1 center, spin 1/2)~\cite{lozovoi_dark_2020}, which shortens the Ramsey coherence time. The experimental results are shown in Fig.~\ref{fig3:Ramsey}(f), where the Ramsey coherence time in the central region is shorter than the outer region, indicating a larger spin density in the central region. In Fig.~\ref{fig3:Ramsey}(g), we show the signal contrast of Ramsey measurement at short times. The region illuminated by the focused laser beam shows a much smaller signal contrast. The contrast decrease under laser illumination is due to the ionization of NV$^-$, which decreases the number of NV centers used for quantum sensing (the NV$^0$ charge state also emits red fluorescence under the green laser, but it cannot be used for quantum sensing due to a different energy structure, so it does not provide any signal contrast).

\section{Discussions and outlook}    
In this work, we report the application of single photon avalanche diode arrays to quantum information processing applications. The SPAD array features fast frame rate, single photon sensitivity, and external gated control, which can be synchronized with state-of-the-art quantum devices to boost their measurement performance. Using a quantum sensor based on solid-state spins as the demonstrating platform, we integrate the SPAD array with a microscope objective to build a fast wide-field quantum imaging setup. With flexible photon collection options, our setup can either be optimized to collect the total fluorescence signal or be tailored to do wide-field quantum imaging with single-photon sensitivity. We use a bulk diamond sample with an NV ensemble to show the capability of our setup in sensing dc and ac magnetic fields, temperature, strain, spin density, and charge dynamics. 

We note that our proof-of-principle demonstration of various quantum sensing applications is based on an NV ensemble sample with $\sim$10~ppm nitrogen density and  $\sim$0.4~ppm NV$^-$ density, which provides a much larger fluorescence intensity than a sample with sparse single NV centers (where the separation of NV is larger than the resolution of the microscope). As a result, we need to strongly attenuate the fluorescence signal to match the photon collection rate of the SPAD camera (at most 1 photon/50~ns, corresponding to 20~M counts/s), and only about 1\%-3\% of the photons collected by the microscope are sent into the camera. This actually allows more than 90\% of the signal to be simultaneously collected by the photodiodes, providing both high sensitivity to the average signal and spatial information on the distribution of the target quantity. With a fast frame rate, our setup is suitable for studying dynamical processes such as the charge transport phenomena in diamond~\cite{lozovoi_optical_2021,jayakumar_optical_2016,wang_manipulating_2023}. Although typical wide-field sensing setups based on CCD-type cameras usually have more pixels and can count larger fluorescence signals, they cannot reach single photon sensitivity and their frame rate is usually ranging from Hz to kHz.  Actually, if both the photon sensitivity and frame rate of the CCD camera can be improved to be similar to our camera, the storage and transmission of the extremely fast-growing data size will be the main challenge in building those devices.
% GW: I will read the related literature to add more citations here

With single-photon sensitivity, our setup is an ideal platform to image an array of single-photon emitters (with a typical separation distance of $\sim1.1\mu$m), which can be created by ion implantation~\cite{wang_high-sensitivity_2015}, or sparse NV ensembles (<100~ppb nitrogen concentration). Using the typical single NV photon collection rate of about 50k counts/s under NA=1.3 to estimate, our setup (NA=0.5) can measure up to around 1000 NV centers within the diffraction-limited detection volume, which corresponds to 25~ppb nitrogen concentration if we assume 100~nm NV layer thickness. Moreover, our setup can be applied to other platforms such as atom arrays trapped in optical tweezers~\cite{saffman_quantum_2010}, which have emerged as a promising system for large-scale quantum computation and quantum simulation. However, one of the major limitations of these platforms is the slow signal readout rate slower than the qubit decoherence, preventing the applications of potential
quantum error correction codes. Integrating our SPAD array with the atom array provides a potential solution to these problems.
%For the first time, an array of SPADs has been used for photon collection from an ensemble of NV centers in diamond, providing \textmu~m-scale spatial resolution and a fast measurement rate. We enhanced the capabilities of NV-based quantum sensors, while at the same time demonstrated a powerful novel application for SPAD arrays. 
%The work included the development of a new optical setup including the SPAD array and the realization of first Microwave field imaging based on Rabi protocol...

%density of diamond is 3.51g/cm$^3$=1.761$\times 10^{23}$/cm$^3$; 1~ppm=1.761$\times 10^{5}$/$\mu$m$^3=1/(17.8\text{nm})^3$.

\acknowledgments
This work was supported in part by DARPA DRINQS program (Cooperative Agreement No. D18AC00024). G.W. thanks  Wenchao Xu, Zeyang Li, Lin Su for help in building the setup. F.M. acknowledges the Rocca program for support and Micro Photon Devices S.r.l. for providing the MPD-SPC3 camera. 

\section*{Author contributions}
G.W. and P.C. conceived the idea. G.W. and F.M. designed and built the setup with assistance from B.L., C.L., and J.X. G.W., F.M., and B.L. wrote the control interface and data processing software. G.W. and F.M. implemented quantum sensing experiments and analyzed the data. All authors discussed the results.

The authors have no conflicts to disclose.

\iffalse
\begin{figure}
\includegraphics[width=0.495\textwidth]{Rabi_1.png}% Here is how to import EPS art
\caption{\label{fig:Rabi_1} A figure caption. The figure captions are
automatically numbered.}
\end{figure}
\begin{figure}
\includegraphics[width=0.495\textwidth]{Rabi_2.png}% Here is how to import EPS art
\caption{\label{fig:Rabi_2} A figure caption. The figure captions are
automatically numbered.}
\end{figure}
\begin{figure}
\includegraphics[width=0.495\textwidth]{Rabi_3.png}% Here is how to import EPS art
\caption{\label{fig:Rabi_3} A figure caption. The figure captions are
automatically numbered.}
\end{figure}

\begin{figure}
\includegraphics[width=0.495\textwidth]{Decoherence_time_1.png}% Here is how to import EPS art
\caption{\label{fig:Decoherence_time_1} A figure caption. The figure captions are
automatically numbered.}
\end{figure}

\begin{figure}
\includegraphics[width=0.495\textwidth]{Decoherence_time_2.png}% Here is how to import EPS art
\caption{\label{fig:Decoherence_time_2} A figure caption. The figure captions are
automatically numbered.}
\end{figure}

\begin{figure}
\includegraphics[width=0.495\textwidth]{Decoherence_time_3.png}% Here is how to import EPS art
\caption{\label{fig:Decoherence_time_3} A figure caption. The figure captions are
automatically numbered.}
\end{figure}
\fi

%\appendix

%\section{Appendixes}

%\nocite{*}
\bibliography{InstrumentPaper}% Produces the bibliography via BibTeX.

\end{document}